\documentclass[11pt, epsfig]{article}
  \usepackage{amsthm,amsfonts}
  \usepackage{amsmath}
\newcommand{\bea}   {\begin{eqnarray}}
\newcommand{\eea}   {\end{eqnarray}}
\begin{document}
\renewcommand{\thefootnote}{\fnsymbol{footnote}}

\thispagestyle{empty}

\title{$D$-module Representations of ${\cal N}=2,4,8$ Superconformal Algebras and Their Superconformal Mechanics }

\author{Zhanna Kuznetsova\thanks{{\em e-mail: zhanna.kuznetsova@ufabc.edu.br}}
~and Francesco
Toppan\thanks{{\em e-mail: toppan@cbpf.br}}
\\
\\
}
\maketitle

\centerline{$^{\ast}${\it UFABC, Rua Catequese 242, Bairro Jardim,}}{\centerline{\it\quad cep 09090-400, Santo Andr\'e (SP), Brazil.}}
{\centerline{
$^{\dag}${\it CBPF, Rua Dr. Xavier Sigaud 150, Urca,}}{\centerline {\it\quad
cep 22290-180, Rio de Janeiro (RJ), Brazil.}}}

\maketitle
\begin{abstract}
The linear (homogeneous and inhomogeneous) $(k, {\cal N}, {\cal N}-k)$ supermultiplets of the ${\cal N}$-extended one-dimensional
Supersymmetry Algebra induce $D$-module representations for the ${\cal N}=2,4,8$ superconformal algebras.\par
 For ${\cal N}=2$, the $D$-module representations
of the $A(1,0)$ superalgebra are obtained. \par
For ${\cal N}=4$ and scaling dimension $\lambda=0$, the
$D$-module representations of the $A(1,1)$ superalgebra are obtained. For $\lambda\neq 0$, the $D$-module representations
of the $D(2,1;\alpha)$ superalgebras are obtained, with $\alpha$ determined in terms of the scaling dimension $\lambda$ according to: $\alpha=-2\lambda$ for $k=4$, i.e. the $(4,4)$ supermultiplet, $\alpha=-\lambda$ for $k=3$, i.e. $(3,4,1)$, and $\alpha=\lambda$ for $k=1$,
i.e. $(1,4,3)$. For $\lambda\neq 0$ the $(2,4,2)$ supermultiplet induces a $D$-module representation for
the centrally extended $sl(2|2)$ superalgebra.\par
For ${\cal N}=8$, the $(8,8)$ root supermultiplet induces a $D$-module representation of the $D(4,1)$
superalgebra at the fixed value $\lambda=\frac{1}{4}$.\par
A Lagrangian framework to construct one-dimensional, off-shell, superconformal invariant actions from single-particle and multi-particles $D$-module representations is discussed. It is applied to explicitly construct invariant actions for the  homogeneous and inhomogeneous ${\cal N}=4$ $(1,4,3)$ $D$-module representations (in the last case for several interacting supermultiplets of different chirality).
\end{abstract}
\vfill

\rightline{CBPF-NF-015/11}

\newpage
\section{Introduction}

In this paper the minimal linear (homogeneous and inhomogeneous) supermultiplets of the
global ${\cal N}$-extended one-dimensional Supersymmetry Algebra
(the dynamical Lie superalgebra of the Supersymmetric Quantum Mechanics \cite{witten})
\bea\label{sqm}
\{Q_i, Q_j \}= 2\delta_{ij} H,&& [H, Q_i]=0,\quad i,j=1,\ldots, {\cal N},
\eea
are used to induce (whenever it is possible) $D$-module representations for Superconformal Algebras.
The constructed $D$-module representations are given in terms of supermatrices whose entries are differential operators in one variable, $t$, which plays the role of the time in physical applications.\par
The paper uses the available classification of the linear global supermultiplets \cite{{pt},
{krt},{top1},{top2},{kt1},{kt2},{gkt},{fg},{dfghil},{dfghil2},{dfghilm},{dfghilm2}}. For ${\cal N}=2,4,8$ the supermultiplets are determined by their $(k, {\cal N}, {\cal N}-k)$ ``field content" \cite{{pt},{krt}}, with
$k=1,2,\ldots, {\cal N}$ (in representation theory the $k={0}$ supermultiplet is identified with the $k={\cal N}$ supermultiplet).  At a given ${\cal N}$, the supermultiplets encode inequivalent mathematical and physical characterizations of the global supersymmetry representations.
For instance, in application to supersymmetric one-dimensional sigma-models \cite{top1}, $k$ denotes the
number of propagating bosons and the dimensionality of their associated target manifold, ${\cal N}$ the number of fermionic degrees of freedom, ${\cal N}-k$
the number of auxiliary bosonic degrees of freedom that are required to close off-shell the superalgebra.\par
The passage from the representations of the global supersymmetry (\ref{sqm}) to the $D$-module representations of a superconformal algebra requires an Ansatz, namely that two extra diagonal even generators $D,K$ have to be added. Together with $H$, the Hamiltonian, they have to close the $sl(2)$ conformal subalgebra \cite{dff}. The scaling (or engineering, see \cite{{krt},{gkt},{dfghilm},{dfghilm2}}) dimension $\lambda$ of the global supermultiplet is now promoted to be the conformal weight of the
$D$-module representation of the superconformal algebra.  The consistency of the procedure requires that
further generators have to be added whenever they arise from the (anti)commutation relations
of the previous generators. After a certain number of steps the procedure can come to a halt.
If this is the case we can identify, for the given $\lambda$, the resulting superalgebra (it satisfies
by construction the graded Jacobi identity). If it turns out to be a superconformal algebra it means that we have ended up with one of its $D$-module representations.\par
The results we obtained can be summarized as follows. \par
For ${\cal N}=2$, the three inequivalent \cite{{pt},{lt}}
minimal linear supermultiplets of the global supersymmetry (the homogeneous $(2,2)$ root supermultiplet, the homogeneous $(1,2,1)$ supermultiplet and the inhomogeneous $(0,2,2)$
supermultiplet \cite{lt}) produce three inequivalent $D$-module representations of the ${\cal N}=2$  $A(1,0)$ superconformal algebra. \par
Much more interesting is the ${\cal N}=4$ case. The ${\cal N}=4$ superconformal algebras 
\cite{{kac},{nrs},{dictionary},{VP}}
are $A(1,1)$, $sl(2|2)$ and the exceptional $D(2,1;\alpha)$ superalgebras which depend on the parameter $\alpha$ (for $\alpha=0,-1$ the $A(1,1)$ superalgebra is recovered). 
For $\lambda=0$, the homogeneous $(k, 4, 4-k)$ global supermultiplets produce $D$-module representations for the $A(1,1)$ ${\cal N}=4$ superconformal algebra. For $\lambda\neq 0$ and $k=1,3,4$, $D$-module  are produced for the $D(2,1;\alpha )$ superconformal algebras with the following identifications between the 
parameter $\alpha$ and the conformal weight $\lambda$:
\bea\label{alphalambda}
&(1,4,3): \alpha = \lambda ;\quad (3,4,1): \alpha=-\lambda; \quad (4,4):  \alpha=-2\lambda.&
\eea
Since $\alpha=-1$ reproduces the $A(1,1)$ superalgebra, for $\lambda\neq 0$ a $D$-module
representation of $A(1,1)$ is recovered for $(1,4,3)$ with conformal weight $\lambda=-1$, for $(3,4,1)$ with conformal weight $\lambda=1$ and for $(4,4)$ with conformal weight $\lambda=\frac{1}{2}$. 
For $k=2$ (the $(2,4,2)$ supermultiplet) we get a different picture for $\lambda\neq 0$. A $D$-module representation for the $sl(2|2)$ ${\cal N}=4$ superconformal algebra is found, with a centrally extended operator which vanishes in the $\lambda\rightarrow 0$ limit.\par
In application to the inhomogeneous global ${\cal N}=4$ supermultiplets \cite{lt} the results are the following. No $D$-module representation for a superconformal algebra is encountered for the inhomogeneous $(0,4,4)$ supermultiplet,  while a $D$-module representation of the $A(1,1)$ superalgebra 
is recovered from the inhomogeneous $(1,4,3)$ supermultiplet (in the inhomogeneous case a consistency condition requires the conformal weight of this supermultiplet to be $\lambda=-1$).\par
For ${\cal N}=8$ we limited our analysis to the ``root" \cite{{pt},{krt},{grt}} supermultiplet $(8,8)$. The result is quite
illuminating. A $D$-module representation of a superconformal algebra is only encountered for a 
single, fixed value of the conformal weight, which turns out to be $\lambda=\frac{1}{4}$. 
There are four superconformal algebras \cite{{kac},{nrs},{dictionary},{VP}} with ${\cal N}=8$ supersymmetry (i.e. $A(3,1)$, $D(4,1)$, $D(2,2)$ and $F(4)$). $D(4,1)$ is the one obtained from our construction.\par
We postpone to the Conclusions the discussion concerning the physical implications of the results here encountered about the construction of $D$-module representations. Here we limit ourselves to point out that the connection between $\alpha$ and $\lambda$ can have  far reaching consequences in model building. To construct superconformal-invariant models for the $D(2,1;\alpha)$ superalgebra with several interacting supermultiplets, their independent conformal weights $\lambda$ have to be adjusted to produce a representation for the given superalgebra. The relation is less trivial than
what  formula (\ref{alphalambda}) seems to suggest. Indeed, due to an $S_3$ symmetry \cite{dictionary}, up to $6$ different values of $\alpha$ produce the same $D(2,1;\alpha)$ superalgebra, see (\ref{equivalpha}). For instance, by specializing $\alpha=-2,-\frac{1}{2}$ or $1$, we obtain the superalgebra $D(2,1)$
belonging to the $D(m,n)$ series. $D$-module representations for $D(2,1)$ are obtained
for $\lambda=1, ~\frac{1}{4}$ or $-\frac{1}{2}$ for the $(4,4)$ supermultiplet, $\lambda= -2, -
\frac{1}{2}$
or $1$ for the $(1,4,3)$ supermultiplet and $\lambda=2,~\frac{1}{2}$ or $-1$ for the $(3,4,1)$ supermultiplet.
One can see the consistency of this analysis with the construction of the $D(4,1)$ $D$-module
representation for the $(8,8)$ supermultiplet at $\lambda=\frac{1}{4}$. From the ${\cal N}=4$ viewpoint
two $(4,4)$ supermultiplets carrying a representation of the $D(2,1)$ subalgebra are linked together by extra generators. Without carrying an ${\cal N}=8$ computation we could have concluded, solely from the ${\cal N}=4$ analysis, that this is indeed possible (necessary but not sufficient condition) only for $\lambda= 1, ~\frac{1}{4}$ or $-\frac{1}{2}$.\par
The next topic we investigated is the construction of off-shell invariant actions for superconformally invariant mechanics \cite{{fr},{ap},{ikt},{fm},{cdkktv},{dipt},{wyl},{bmsv},{pap}}. The crucial ingredient is the information contained in the $D$-module representations which naturally leads to Lagrangian systems. The construction makes use of the techniques (see \cite{{krt},{grt},{gkt}}) developed to build off-shell invariant actions for global supersymmetries by applying supercharges (acting as graded Leibniz derivative) on given prepotentials. The implementation of the full superconformal invariance requires extra conditions
to be satisfied. The extra-conditions enter as constraints for the prepotentials. The scale-invariance of the resulting action is a particular example of an extra condition required from the dilatation invariance.  We formulate the general problem and, to be specific, we applied it to the concrete cases of the ${\cal N}=4$ homogeneous $(1,4,3)$ supermultiplet (for arbitrary values of $\lambda$) and the inhomogeneous $(1,4,3)$ supermultiplet (the general case, with $n_+$ supermultiplets of positive chirality and $n_-$ supermultiplets of negative chiralities, see \cite{{top3},{gh}}). The inhomogeneous supermultiplets are quite important. Indeed, the inhomogeneity is essential to produce in the
superconformal action terms associated with the Calogero potentials \cite{{gal},{bgl},{glp},{bks}}.
The supertransformations that  we can interpret \cite{lt} as inhomogeneous global supermultiplets were originally derived from special constraints on given superfields \cite{{ikp},{dprt}}.\par
This is the right point to discuss the relation between our paper and other works on superconformal mechanics. Superconformal mechanic is a vast and active field of research with applications 
to the dynamics of test particles in proximity of the horizon of certain black holes \cite{{bmsv},{bkss}}.
By using the superfield approach superconformal mechanical systems have been constructed  for ${\cal N}=2$ \cite{gal2}, for ${\cal N}=4$ based on $A(1,1)$ \cite{{dprst},{bko}} and $D(2,1;\alpha)$ \cite{{ikl},{ikl0},{in},{fil},{fil2},{kl}}, for ${\cal N}=8$  \cite{{bikl1},{di}}. The superconformal invariance has been realized in Hamiltonian framework \cite{bgk} or in terms of a Poisson brackets structure, non-linearly \cite{acp} and so on. Multiparticle superconformal systems have been investigated \cite{gl0}. From the perspective of the supersymmetry representation no new ingredient enters with respect to the single-particle superconformal mechanics; their construction, however, requires the challenging and still open mathematical problem of finding solutions to the WDVV equations \cite{glp}.
\par
The drawbacks of the superspace approach are known. For large ${\cal N}$ to get irreducible representations the superfields have to be constrained. For global supersymmetry the analysis of the minimal supermultiplets \cite{{pt},{krt},{kt1},{kt2}} gives the general linear solution. It can also  be phrased as finding the admissible constraints that have to be put on a general, unconstrained, superfield
\cite{{dfghilm},{dfghilm2}}. \par
The classification of the minimal supermultiplets is mandatory to get a systematic construction
of supersymmetric models. It requires, as we have mentioned, an approach to the construction of supersymmetric-invariant off-shell actions which is distinct from the one given by superspace. In application to global supersymmetry this approach proved useful in determining one-dimensional supersymmetric sigma-models which had not been previously identified by using the superfield techniques \cite{krt}. The present work extends the systematic
approach from minimal global supermultiplets to the arena of superconformal algebra. The discovered relation between the conformal weight $\lambda$ of the $D$-module and its associated superconformal algebra is very neat. It is the basis to start a systematic investigation of the superconformal mechanics in a Lagrangian framework. \par
The scheme of the paper is as follows. In Section {\bf 2} we briefly review, for completeness, the 
classification of the finite ${\cal N}=2,4,8$ superconformal Lie algebras. In Section {\bf 3}
we summarize the main results concerning the classification of minimal linear supermultiplets
(both homogeneous and inhomogeneous) for global supersymmetry and outline their extension to $D$-module representations for superconformal algebras. In Section {\bf 4} we present the $D$-module
representations for the ${\cal N}=2$ superconformal algebra $A(1,1)$. In Section {\bf 5} we present the results of our investigation of $D$-module representations for ${\cal N}=4$ superconformal algebras for the full list of minimal linear ${\cal N}=4$ supermultiplets. In Section {\bf 6} we present 
the $D$-module representation for the $D(4,1)$ superalgebra obtained from the ${\cal N}=8$ $(8,8)$ root supermultiplet at the fixed value $\lambda=\frac{1}{4}$ of the conformal weight. The construction of superconformal-invariant mechanics, based on $D$-module representations
of superconformal algebras, is outlined in Section {\bf 7}. As an example, the construction of ${\cal N}=4$ superconformally invariant actions, both for the homogeneous and inhomogeneous $(1,4,3)$ supermultiplets, is
explicitly carried  out. In the Conclusions we comment about our results, pointing out their relevance for the construction of superconformally invariant mechanical systems. We will compare our framework with several different approaches and constructions found in the literature. We will also sketch which research topics could profit from the present investigation. 

\section{The ${\cal N}=2,4,8$ Superconformal Algebras}

We present for completeness the list of ${\cal N}=2,4,8$ Superconformal Algebras in one dimension. They are recovered from the basic Lie superalgebras \cite{{kac},
{nrs}, {dictionary},{VP}} and satisfy the following conditions: the total number
of odd generators is $2 {\cal N}$, while the even sector subalgebra is given by a direct sum
$sl(2)\oplus R$. The $sl(2)$ subalgebra  is known as the conformal bosonic algebra, while $R$ is known as $R$-symmetry. We follow the conventions of \cite{dictionary} and present them as complex Lie superalgebras (they possess different real forms).
\par
The ${\cal N}=2$ S.C.A. is $A(1,0)$, with $4$ even and $4$ odd generators. The $R$-symmetry is $u(1)$. $A(1,0)$ is often indicated as $sl(2|1)$.\par
The ${\cal N}=4$ S.C.A.'s are:\\
{\em i}) the $A(1,1)$ superalgebra ($A(1,1)=sl(2|2)/{\cal Z})$, with $6$ even generators and bosonic
sector given by $sl(2)\oplus sl(2)$;\\
{\em ii}) the $sl(2|2)$ superalgebra, possessing $7$ even generators and bosonic sector given by $sl(2)\oplus
sl(2)\oplus u(1)$;\\
{\em iii}) the exceptional Lie superalgebras $D(2,1;\alpha)$,  depending on the complex parameter
$\alpha\neq 0,-1$ (for these values and in the limit $\alpha\rightarrow\infty$ the $A(1,1)$
superalgebra is recovered), possessing $9$ even generators and bosonic sector given by
$sl(2)\oplus sl(2)\oplus sl(2)$.\\
The superalgebras obtained by values of $\alpha$ related by the $S_3$ symmetry
generated by $\alpha\mapsto -(1+\alpha)$, $\alpha\mapsto \frac{1}{\alpha}$, namely
\bea\label{equivalpha}
&\alpha, ~{1}/{\alpha}, ~ -(1+\alpha),~ -{1}/({1+\alpha}), ~-{(1+\alpha)}/{\alpha},~
-{\alpha}/({1+\alpha}),&
\eea  
are isomorphic.\par
For $\alpha$ real the following fundamental domains for $D(2,1;\alpha)$  can be chosen:
$1\leq \alpha <\infty$, $0<\alpha\leq 1$, $-\frac{1}{2}\leq \alpha <0$, $-1<\alpha\leq -\frac{1}{2}$, 
$-2\leq \alpha <-1$ or $-\infty<\alpha \leq -2$.\par
The special (``boundary") values $\alpha=-2,-\frac{1}{2},1$ correspond to the $D(2,1)$
superalgebra which belongs to the $D(m,n)=osp(2m|2n)$ series.\par
The ${\cal N}=8$ S.C.A.'s with $16$ odd generators are:
\\
{\em i}) the $A(3,1)=sl(4|2)$ superalgebra, possessing  $19$ even generators and bosonic sector given by $sl(2)\oplus sl(4)\oplus u(1)$,  \\
{\em ii}) the $D(4,1)=osp(8,2)$ superalgebra, possessing $31$ even generators and bosonic sector given by $sl(2)\oplus so(8)$,\\
{\em iii}) the $D(2,2)=osp(4|4)$ superalgebra, possessing $16$ even generators and bosonic
sector given by $sl(2)\oplus so(3)\oplus sp(4)$,\\
{\em iv}) the $F(4)$ exceptional superalgebra, possessing $24$ even generators and bosonic sector given by $sl(2)\oplus so(7)$.\par
In the following we will identify which global supermultiplet induces a $D$-module representation for which superconformal algebra. 
The ${\cal N}=4$ case is particularly intricate due to the presence of the $\alpha$ parameter entering $D(2,1;\alpha)$. The parameter $\alpha$ (and, as a consequence, the associated superconformal algebra) is determined in terms of the scaling (or engineering) dimension
of the given initial supermultiplet.

\section{$D$-module representations of superconformal algebras}

We summarize here for convenience the main results, ingredients and conventions \cite{{pt},{krt},{kt1},{kt2},{gkt}} concerning the classification of minimal linear supermultiplets for the global superalgebra (\ref{sqm}) with ${\cal N}=2,4,8$.\par
The minimal supermultiplets contain ${\cal N}$ bosonic and ${\cal N}$ fermionic fields. Inequivalent supermultiplets ($x_i(t);\psi_a(t); g_m(t))$ are characterized by their ``field content" $(k, {\cal N}, {\cal N}-k)$ ($i=1,\ldots, k$, $a =1,\ldots {\cal N}$, $m=1,\ldots {\cal N}-k$). The component fields are real functions of $t$.  For our purposes here we can assume  the $x_i$ and $g_m$ fields to be bosonic and the $\psi_a$ fields to be fermionic. Due to their relation with one-dimensional supersymmetric sigma-models the $x_i$ fields will be called {\em propagating bosons} or {\em target coordinates}, the $g_m$ fields will be called {\em auxiliary fields}. 
A scale-dimension (also known as ``engineering dimension" or ``mass-dimension") is assigned to the component fields. The assignment is given by  $[x_i]=\lambda$, $[\psi_a]=\lambda+\frac{1}{2}$, $[g_m]=\lambda+1$. In representation theory the parameter
$\lambda$ is arbitrary and can be left indeterminate. The different scale-dimensions for component fields is in order to make them compatible with the scale-dimension assignment $[t]=-1$ for the time coordinate, $[Q_i]=\frac{1}{2}$ and $[H]=1$ for the operators entering (\ref{sqm}). $H\equiv {\bf 1}\cdot\frac{d}{dt}$  will be called the {\em Hamiltonian operator}, while the $Q_i$'s are called {\em global supercharges}.\par
The $({\cal N}, {\cal N})$ supermultiplets (for ${\cal N}=2,4,8$) are known as {\em root supermultiplets}. They are uniquely determined \cite{pt} by their associated Clifford algebra representation. 
Their supertransformations can be encoded in $2{\cal N}\times 2{\cal N}$ supermatrices whose entries are differential operators (the non-vanishing entries are either $\pm1$ or $\pm \frac{d}{dt}$).
The remaining
supermultiplets (for $k<{\cal N}$) are obtained from the given root supermultiplet via a dressing transformation determined by the dressing operator $S_k$. $S_k$ is a diagonal operator with $k$ entries $\partial_t=\frac{d}{dt}$ in the upper-left block (its remaining $2{\cal N}-k$ diagonal entries are $1$).  \par
Let $Q_i^R$ be the supermatrices expressing the supercharges in the root representation. The supermatrices $Q_i^{k}$ of a dressed representation are given by
\bea\label{qdressing}
Q_i^{k} &=& S_kQ_i^RS_k^{-1}.
\eea
Even if $S_k^{-1}$ is a pseudo-differential operator, the supermatrices $Q_i^{k}$ are differential operators (see the analysis in \cite{{pt},{krt}}). 
\par
We conclude this quick review on global supersymmetry by pointing out that, for ${\cal N}=4$,
the supermultiplets are associated with a chirality $\pm 1$. The chirality is due to the relation between the ${\cal N}=4$ root supermultiplet and the quaternions. It results in the choice of the overall sign 
for the totally antisymmetric tensor $\epsilon_{ijk}$ ($\epsilon_{123}=\pm1$). For ${\cal N}=8$
a similar chiral relation exists between the root supermultiplet and the octonionic structure constants $C_{ijk}$ \cite{{krt},{top3}}. A mirror symmetry flips the chirality of a {\em single} supermultiplet. On the other hand, it simultaneously flips the chirality of $n$ different ${\cal N}=4$ supermultiplets. As a consequence, inequivalent ${\cal N}=4$ off-shell actions involving  $n_\pm$ supermultiplets of given chirality (the total number being  $n=n_++n_-$) are determined by the
modulus $|n_+-n_-|$. In Section {\bf 7} we derive such a consequence for $(1,4,3)$ $ {\cal N}=4$
supermultiplets.\par
In \cite{lt} it has been pointed out that inhomogeneous supermultiplets can be consistently defined for the global (\ref{sqm}) superalgebra. This situation arises when the auxiliary fields have
$0$ scale-dimension. In this case the supersymmetry transformations of the fermions can produce not only auxiliary fields and time-derivatives of the propagating bosons, but also real constants. A real inhomogeneity appears when at least one constant cannot be reabsorbed via
field redefinitions. The presence of the inhomogeneity drastically reduces the admissible supermultiplets. The closure of the (\ref{sqm}) supersymmetry algebra implies that for
${\cal N}=2$ a inhomogeneous term can be added to the $(0,2,2)$ supermultiplet (but not to the
$(1,2,1)$ supermultiplet). For ${\cal N}=4$, inhomogeneous supermultiplets are encountered
for $(0,4,4)$ and $(1,4,3)$; for ${\cal N}=8$ they are encountered for $(0,8,8)$, $(1,8,7)$, $(2,8,6)$ and $(3,8,5)$. Stated otherwise,  for those values of field content the homogeneous supermultiplets can be extended to accommodate inhomogeneous terms that cannot be reabsorbed through field redefinitions.\par
The inhomogeneous supermultiplets can also be represented in terms of supermatrices
(whose entries are differential operators in $t$). The constant can be added as an extra entry
in the supermultiplet. In the ${\cal N}=2$ $(0,2,2)$ case, for instance, the supermatrices 
act on the supermultiplet $|m>$, whose dual is $|m>^T=<m|=(\psi_1,\psi_2;g_1,g_2,1)$.
The supercharges are $5\times 5$ supermatrices (whose ${\bf Z}_2$-grading decomposition is $5=(2|3)$). Similarly, in the ${\cal N}=4$ case, the inhomogeneous supermultiplet $(1,4,3)$
is expressed by $9\times 9$ supermatrices ($9=(5|4)$ in ${\bf Z}_2$-grading) acting on the
supermultiplet $|m>$, where now $|m>^T =<m|=(x, g_1,g_2,g_3,1;\psi_1,\psi_2,\psi_3,\psi_4)$. In the following we will give the explicit presentation of the supermatrices associated to the inhomogeneous supermultiplets. An important observation is the following. Since the constant $1$
enters the $l^{th}$ row of the supermultiplet $|m>$ (e.g., $l=5$ in the above inhomogeneous $(1,4,3)$
example), then  the supermatrix entries $M_{il}$ proportional to $\partial_t$ (or a higher derivative) have no effect in the action on $|m>$ (due to the left action $\partial_t (1)=0$). This property has to be taken into account when computing (anti)commutators to check the closure of the supermatrix algebra. On the other hand, supermatrices are just a convenient tool to deal with the supertranformations of the component fields; the closure of the superalgebra can be independently checked from the supertransformations alone (without presenting them in supermatrix forms). The two  pictures obviously agree producing the same results.\par
We discuss now the extension from supermultiplets of global supersymmetry to $D$-module representations of superconformal algebras. As anticipated in the Introduction, the strategy consists in introducing two extra diagonal generators $D,K$ which, together with $H$, close an $sl(2)$ conformal algebra. $D,K$ act on a component field $\varphi$ of weight $\lambda'$ according to
\bea\label{daction}
D\varphi=-t\partial_t\varphi-\lambda' \varphi, &&K\varphi=-t^2\partial_t\phi -2\lambda't\varphi.
\eea
The consistency of the procedure requires the identification of the weight $\lambda'$ with the scale-dimension of the corresponding component field entering the supermultiplet  (otherwise the commutators
$[D,Q_i]=\frac{1}{2}Q_i$, producing the conformal weight of the supercharges, cannot be recovered).  Therefore, the only free parameter in the construction is the scale-dimension $\lambda$ of the propagating bosons. $\lambda$ will be identified with the overall conformal weight of the supermultiplet. The superconformal partners ${\widetilde Q}_i$ are produced from the commutators $[K,Q_i]={\widetilde Q}_i$. Extra generators have to be added as already explained.\par
With respect to the global supersymmetry a key difference is the fact that now the dressing transformation
\bea\label{ddressing}
D&\mapsto& D^k=S_k DS_k^{-1}
\eea
produces a differential operator $D^k$ only for $\lambda=0$. For $\lambda=0$ superconformal algebras are produced from the ${\cal N}=2$ and ${\cal N}=4$ root supermultiplets. All dressed
generators are differential operators so that, at $\lambda=0$, the $(k,{\cal N}, {\cal N}-k)$ dressed supermultiplets automatically give a $D$-module representation for the same superconformal algebra. For $\lambda\neq 0$ the situation is different. The absence of the dressing explains why
the ${\cal N}=4$ $(k, 4, 4-k)$ superconformal multiplets produce $D$-module 
representations for different
superconformal algebras at a given $\lambda\neq 0$.\par
This approach to construct $D$-module representations of superconformal algebras is based on two ingredients:\\
{\em i}) a minimal linear supermultiplet (either homogeneous or inhomogeneous) of the global supersymmetry taken as a starting point and\\
{\em ii}) the assignment of an overall scale-dimension $\lambda$ to the supermultiplet.
\par
An automatic procedure is then started which may eventually end up with a $D$-module representation of a
superconformal algebra (to be identified).\par
This construction was first used in \cite{bt} in a very specific case (the determination of the $A(1,0)$ superconformal algebra from the ${\cal N}=2$ $(1,2,1)$ supermultiplet).\par
In Sections {\bf 4} and {\bf 5} we present the results of our  investigation for the full list
of linear supermultiplets of
the ${\cal N}=2$ and ${\cal N}=4$ superconformal algebras, respectively.

\section{$D$-module reps of the $A(1,0)$ ${\cal N}=2$ SCA}

For our purposes the $A(1,0)$ superconformal algebra can be presented in terms of the even generators $H,D,K,W$ and the odd generators $Q_i, {\widetilde Q}_i$ ($i=1,2$). 
The $sl(2)$ conformal subalgebra is generated by $H,D,K$, with $D$ the dilatation operator,
while $H$ plays the role of the Hamiltonian. $Q_1,Q_2$ are the global supercharges
and ${\widetilde Q}_1, {\widetilde Q}_2$ their superconformal partners. The $u(1)$ generator is $W$. Explicitly, the non-vanishing (anti)commutation relations are
\bea\label{a10}
&
\begin{array}{lllllllll}
\relax [D,H]&=&H,~&[D,K]&=&-K,~& [H,K]&=&2D,\\
\relax [D,Q_i]&=&\frac{1}{2} Q_i,~&[D,{\widetilde Q}_i]&=&-\frac{1}{2}{\widetilde Q}_i,~&&& \\
\relax [H,{\widetilde Q}_i]&=&Q_i,~& [K,{Q}_i]&=&{\widetilde Q}_i ,~&&& \\
\relax [W, Q_i]&=&-\epsilon_{ij}Q_j,~& [W,{\widetilde Q}_i]&=&-\epsilon_{ij}{\widetilde Q}_{ij},~&&& \\
\{Q_i,Q_j\}&=&2\delta_{ij}H, ~& \{{\widetilde Q}_i, {\widetilde Q}_j\}&=&-2\delta_{ij}K,~&&& \\
\{Q_i,{\widetilde Q}_j\}&=&-2\delta_{ij}D+\epsilon_{ij}W,~&&&&&&
\end{array}
&
\eea
with $\epsilon_{12}=-\epsilon_{21}=1$.\par
In the subsections below we present its $D$-module representations based on, respectively, the following global ${\cal N}=2$ supermultiplets:
\\
{\em i}) the homogeneous $(2,2)$ supermultiplet,
\\
{\em ii})  the homogeneous $(1,2,1)$ supermultiplet and\\
{\em iii}) the inhomogeneous $(0,2,2)$ supermultiplet.
\par
The closure of the superalgebra (\ref{a10}) and, in particular, of the commutator $[D,X]= xX$, 
where $x$ is the scaling dimension of the operator $X$, requires that the conformal weights
of the component fields entering the $(k,{\cal N}, {\cal N}-k)$ supermultiplet have to be given by
$(\lambda, \lambda+\frac{1}{2},\lambda+1)$, respectively. In the inhomogeneous case the  consistency condition further restricts the auxiliary component fields to possess the same scaling dimension $(=0)$ of a real constant. These requirements hold for the ${\cal N}=4,8$ superconformal algebras as well.

\subsection{The $(2,2)$ ``root" representation}

The $D$-module representation based on the homogeneous $(2,2)$ ``root" supermultiplet
$|m>$ (such that $<m|^T=(x_1,x_2;\psi_1,\psi_2)$ ), expressed in terms
of $4\times 4$ supermatrices whose entries are $t$-dependent differential operators, 
is given by (here and in the rest of the paper $E_{ij}$ denotes the supermatrix with entry $1$ at the crossing of the $i^{th}$ row and $j^{th}$ column and $0$ otherwise)
\bea\label{22}
H&=&(E_{11}+E_{22}+E_{33}+E_{44})\partial_t,\nonumber\\
D&=&-(E_{11}+E_{22}+E_{33}+E_{44})(t\partial_t+\lambda) -\frac{1}{2}(E_{33}+E_{44}),\nonumber\\
K&=&-(E_{11}+E_{22}+E_{33}+E_{44})(t^2\partial_t+2\lambda t) -(E_{33}+E_{44})t,
\nonumber\\
W&=&E_{34}-E_{43} -2\lambda (E_{12}-E_{21}+E_{34}-E_{43}), \nonumber\\
Q_1&=&E_{13}+E_{24}+(E_{31}+E_{42})\partial_t,\nonumber\\
Q_2&=&E_{14}-E_{23}-(E_{32}-E_{41})\partial_t,\nonumber\\
{\widetilde Q}_1 &=& (E_{13}+E_{24})t+(E_{31}+E_{42})(t\partial_t+2\lambda ),\nonumber\\
{\widetilde Q}_2 &=& (E_{14}-E_{23})t-(E_{32}-E_{41})(t\partial_t+2\lambda ).
\eea

\subsection{The $(1,2,1)$ representation}

The $D$-module representation based on the homogeneous $(1,2,1)$ supermultiplet
$|m>$ (such that $<m|^T=(x, g;\psi_1,\psi_2)$ )  is given \cite{{pt},{krt}} by
\bea\label{121}
H&=&(E_{11}+E_{22}+E_{33}+E_{44})\partial_t,\nonumber\\
D&=&-(E_{11}+E_{22}+E_{33}+E_{44})(t\partial_t+\lambda) -\frac{1}{2}(2E_{22}+E_{33}+E_{44}),\nonumber\\
K&=&-(E_{11}+E_{22}+E_{33}+E_{44})(t^2\partial_t+2\lambda t) -(2E_{22}+E_{33}+E_{44})t,
\nonumber\\
W&=&E_{34}-E_{43} -2\lambda (E_{12}-E_{21}+E_{34}-E_{43}), \nonumber\\
Q_1&=&E_{13}+E_{42}+(E_{31}+E_{24})\partial_t,\nonumber\\
Q_2&=&E_{14}-E_{32}-(E_{23}-E_{41})\partial_t,\nonumber\\
{\widetilde Q}_1 &=& (E_{13}+E_{42})t+(E_{31}+E_{24})(t\partial_t+2\lambda )+E_{24},\nonumber\\
{\widetilde Q}_2 &=& (E_{14}-E_{32})t-(E_{23}-E_{41})(t\partial_t+2\lambda )-E_{23}.
\eea
\subsection{The inhomogeneous $(0,2,2)$ representation}

The $D$-module representation based on the inhomogeneous $(0,2,2)$ supermultiplet
$|m>$ (such that $<m|^T=(\psi_1,\psi_2; g_1,g_2, 1)$ ) is expressed in terms of the left action 
on $|m>$ resulting from the $5\times 5$ supermatrices (with $5=(2|3)$ in the  ${\bf Z}_2$-grading
decomposition). A rotation of the $Q_1, Q_2$ global supersymmetry plan can leave us, without loss of generality, with a single inhomogeneous parameter $c$. In the presentation below $c$ only enters the $Q_2$, ${\widetilde Q}_2$ and $W$ transformations. The $D$-module representation of the homogeneous root supermultiplet at $\lambda=-\frac{1}{2}$ is recovered in the limit $c\rightarrow 0$. We have 
\bea\label{022in}
H&=&(E_{11}+E_{22}+E_{33}+E_{44})\partial_t,\nonumber\\
D&=&-(E_{11}+E_{22}+E_{33}+E_{44})(t\partial_t) +\frac{1}{2}(E_{11}+E_{22}),\nonumber\\
K&=&-(E_{11}+E_{22}+E_{33}+E_{44})(t^2\partial_t) +(E_{11}+E_{22})t,
\nonumber\\
W&=&E_{12}-E_{21} +2E_{34}-2E_{43}+cE_{45}, \nonumber\\
Q_1&=&E_{13}+E_{24}+(E_{31}+E_{42})\partial_t,\nonumber\\
Q_2&=&E_{14}-E_{23}-(E_{32}-E_{41})\partial_t+cE_{25},\nonumber\\
{\widetilde Q}_1 &=& (E_{13}+E_{24})t+(E_{31}+E_{42})(t\partial_t-1 ),\nonumber\\
{\widetilde Q}_2 &=& (E_{14}-E_{23})t-(E_{32}-E_{41})(t\partial_t-1)+tcE_{25}.
\eea

\section{$D$-module reps of the $A(1,1)$ and $D(2,1;\alpha)$ ${\cal N}=4$ SCA's}

We present the superalgebra $D(2,1;\alpha)$ in terms of the $4$ supercharges $Q_I$, their superconformal partners ${\widetilde Q}_I$
($I=1,2,3,4$), the even generators $H,D,K$ (closing the $sl(2)$ conformal subalgebra), 
$S_i$ and $W_i$ ($i=1,2,3$). The superalgebra $A(1,1)$ is recovered for $\alpha=0$ by disregarding the $W_i$ generators (alternatively, it can be recovered from $\alpha=-1$
and the consistent constraint $S_i+W_i=0$). The non-vanishing (anti)commutation relations are given by
\bea\label{d21alpha}
&
\begin{array}{llllll}
[H,K]&=&2D,~&&&\\
\relax [D,H]&=&H,~&[D,K]&=&-K, \\
\relax [D,Q_I]&=&\frac{1}{2} Q_I,~&[D,{\widetilde Q}_I]&=&-\frac{1}{2}{\widetilde Q}_i,\\
\relax [H,{\widetilde Q}_I]&=&Q_I,~& [K,{Q}_I]&=&{\widetilde Q}_I , \\
\{Q_I,Q_J\}&=&2\delta_{IJ}H, ~& \{{\widetilde Q}_I, {\widetilde Q}_J\}&=&-2\delta_{IJ}K, \\
\{Q_4,{\widetilde Q}_4\}&=&-2D~,&\{Q_i,{\widetilde Q}_j\}&=&-2\delta_{ij}D-\epsilon_{ijk}(S_k-2\alpha W_k),\\
\{Q_4,{\widetilde Q}_i\}&=&S_i,~&\{{\widetilde Q}_4,{ Q}_i\}&=&-S_i,\\
\relax [S_i, Q_4]&=&-Q_i,~&[S_i,Q_j]&=&\delta_{ij}Q_4+\epsilon_{ijk}(1+2\alpha)Q_k,\\
\relax [S_i, {\widetilde Q}_4]&=&-{\widetilde Q}_i,~&[S_i,{\widetilde Q}_j]&=&\delta_{ij}{\widetilde Q}_4+\epsilon_{ijk}(1+2\alpha){\widetilde Q}_k,\\
\relax [W_i, Q_4]&=&Q_i,~&[W_i,Q_j]&=&-\delta_{ij}Q_4+\epsilon_{ijk}Q_k,\\
\relax [W_i, {\widetilde Q}_4]&=&{\widetilde Q}_i,~&[W_i,{\widetilde Q}_j]&=&-\delta_{ij}{\widetilde Q}_4+\epsilon_{ijk}{\widetilde Q}_k,\\
\relax [S_i,S_j]&=&\epsilon_{ijk}(2(1+\alpha)S_k-2\alpha W_k),~
& [S_i, W_j]&=&\epsilon_{ijk} 2\alpha W_k, \\
\relax [W_i,W_j]&=&2\epsilon_{ijk}W_k,~&&&
\end{array}
&
\eea
where $\epsilon_{ijk}$ is the totally antisymmetric tensor ($\epsilon_{123}=1$).

\subsection{The $(4,4)$ root rep with $\lambda$ scaling dimension}

For scaling dimension $\lambda=0$ the homogeneous root supermultiplet of the global ${\cal N}=4$-extended one-dimensional supersymmetry induces the $(4,4)$ $D$-module representation of the $A(1,1)$ superalgebra.
\par
A presentation is given by the supermatrices
\bea\label{44}
&&
\begin{array}{ll}
Q_i= \left(\begin{array}{cc}
0\quad&R_i\\
-R_i\partial_t\quad&0
\end{array}\right), & Q_4=\left(\begin{array}{cc}
0\quad&{\bf 1}_4\\
{\bf 1}_4\partial_t\quad&0
\end{array}\right),\\
{\widetilde Q}_i =tQ_i,& {\widetilde Q}_4=tQ_4,\\
H= {\bf 1}_8\cdot\partial_t, & S_i=\left(\begin{array}{cc}
0&0\\
0&R_i
\end{array}\right),\\
D=-{\bf 1}_8\cdot t\partial_t-\frac{1}{2}Y, & 
K=-{\bf 1}_8 \cdot t^2\partial_t-tY,
\end{array}
\eea
where, without loss of generality, the $R_i$'s can be chosen
{\small \bea\label{ri}
&R_1=\left(\begin{array}{cccc}
0&1&0&0\\
-1&0&0&0\\0&0&0&-1\\
0&0&1&0
\end{array}\right),\quad
R_2=\left(\begin{array}{cccc}
0&0&1&0\\
0&0&0&1\\-1&0&0&0\\
0&-1&0&0
\end{array}\right),\quad R_3=\left(\begin{array}{cccc}
0&0&0&1\\
0&0&-1&0\\0&1&0&0\\
-1&0&0&0
\end{array}\right),&
\eea
}
so that $R_iR_j=-\delta_{ij}+\epsilon_{ijk}R_k$. The supermatrix $Y$ is
\bea\label{y}
Y&=&\left(\begin{array}{cc}
0&0\\
0&{\bf 1}_4
\end{array}\right).
\eea
For $\lambda \neq 0$, the $(4,4)$ $D$-module representation of the $D(2,1;\alpha)$ 
superalgebra is obtained.
The relation between the parameter $\alpha$ and the scaling dimension $\lambda$ is expressed by
\bea\label{alphalambda44}
\alpha &=& -2\lambda.
\eea
 The $D$-module generators for $\lambda\neq 0$ are
\bea\label{44lambda}
&
\begin{array}{lll}
H^{\lambda}=H,&D^\lambda= D-\lambda {\bf 1}_8 ,& K^\lambda= K-2\lambda t {\bf 1}_8,\\
Q^\lambda_i=Q_i,& Q^\lambda_4=Q_4,&~ \\
{\widetilde Q}^\lambda_i={\widetilde Q}_i-2\lambda 
\left(\begin{array}{cc}
0&0\\
R_i&0
\end{array}\right)
,&{\widetilde Q}^\lambda_4=  {\widetilde Q}_4+2\lambda \left(\begin{array}{cc}
0&0\\
{\bf 1}_4&0
\end{array}\right),&~ \\
S^\lambda_i=S_i-2\lambda \left(\begin{array}{cc}
R_i&0\\
0&R_i
\end{array}\right), & W^\lambda_i= 
\left(\begin{array}{cc}
R_i&0\\
0&0
\end{array}\right)
. &~ 
\end{array}
&
\eea

\subsection{The $(1,4,3)$ rep with $\lambda$ scaling dimension}

For $\lambda=0$, the $(1,4,3)$ $D$-module representation of the $A(1,1)$ superalgebra  is obtained by dressing
the $\lambda=0$ $D$-module representation of the root supermultiplet $(4,4)$. The dressing
operator is the diagonal $8\times 8$ matrix $S$, with diagonal elements 
$diag (S)=(1,\partial_t,\partial_t,\partial_t,1,1,1,1)$. Let $X$ be a generator of the $\lambda=0$
root $D$-module representation of $A(1,1)$. The corresponding generator ${\overline X}$ of the $(1,4,3)$
$D$-module representation is given by $
{\overline X}= S X S^{-1}$ (in order to avoid burdening unnecessarily the notation, in the remaining subsections the overline will be used for a different set of dressed operators). \par
For $\lambda\neq 0$ the homogeneous $(1,4,3)$ $D$-module representation of the $D(2,1;\alpha)$ superalgebra is obtained. It is expressed by the supermatrices $X^\lambda$. The relation between $\alpha$ and $\lambda$ is given by
\bea\label{alphalambda143}
\alpha&=& \lambda.
\eea
We have
\bea\label{143}
H^\lambda &=&{\overline H},\nonumber\\
D^\lambda &=& {\overline D}-\lambda {\bf 1}_8, \nonumber\\
K^\lambda &=& {\overline K}-2\lambda t {\bf 1}_8, \nonumber\\
Q_I^\lambda &=&{\overline Q}_I,\quad\quad\quad\quad \quad\quad\quad\quad\quad\quad\quad\quad \quad\quad\quad\quad\quad\quad\quad\quad (I=1,2,3,4),\nonumber\\
{\widetilde Q}^\lambda_1&=&{\overline{\widetilde Q}}_1+2\lambda
(-E_{25}-E_{38}+E_{47}+E_{61}), \nonumber\\
{\widetilde Q}^\lambda_2&=&{\overline{\widetilde Q}}_2+2\lambda
(E_{28}-E_{35}-E_{46}+E_{71}), \nonumber\\
{\widetilde Q}^\lambda_3&=&{\overline{\widetilde Q}}_3+2\lambda
(-E_{27}+E_{36}-E_{45}+E_{81}), \nonumber\\
{\widetilde Q}^\lambda_4&=&{\overline{\widetilde Q}}_4+2\lambda
(E_{26}+E_{37}+E_{48}+E_{51}), \nonumber\\
S^\lambda_1&=&{\overline S}_1+2\lambda (-E_{34}+E_{43}-E_{78}+E_{87}) ,\nonumber\\
S^\lambda_2&=&{\overline S}_2+2\lambda (E_{24}-E_{42}+E_{68}-E_{86}),\nonumber\\
S^\lambda_3&=&{\overline S}_3+2\lambda (-E_{32}+E_{23}-E_{67}+E_{76}),\nonumber\\
W^\lambda_1&=&{\overline W}_1+2\lambda (-2E_{34}+2E_{43}-E_{56}+E_{65}-E_{78}+E_{87}),\nonumber\\
W^\lambda_2&=&{\overline W}_2+2\lambda (2E_{24}-2E_{42}-E_{57}+E_{68}+E_{75}-E_{86}),\nonumber\\
W^\lambda_3&=&{\overline W}_3+2\lambda (-2E_{23}+2E_{32}-E_{58}-E_{67}+E_{76}+E_{85}).
\eea

\subsection{The $(2,4,2)$ rep with $\lambda$ scaling dimension}

For $\lambda=0$, the $(2,4,2)$ $D$-module representation of the $A(1,1)$ superalgebra  is obtained by dressing
the $\lambda=0$ $D$-module representation of the root supermultiplet $(4,4)$. The dressing
operator is the diagonal $8\times 8$ matrix $S$, with diagonal elements 
$diag (S)=(1,1,\partial_t,\partial_t,1,1,1,1)$. Let $X$ be a generator of the $\lambda=0$
root $D$-module representation of $A(1,1)$. The corresponding generator ${\overline X}$ of the $(2,4,2)$
$D$-module representation is given by $
{\overline X}= S X S^{-1}$. 
\par
For $\lambda\neq 0$ one obtains in this case (unlike the other $(k,4,4-k)$ supermultiplets with $k\neq 2$) a $D$-module representation of the $sl(2|2)$ superalgebra (whose generators are denoted as ``$X^\lambda$") containing, with respect to $A(1,1)$, an extra
generator (the central extension $F^\lambda$).  \par
The $sl(2|2)$ $D$-module representation is given by
\bea\label{242}
H^\lambda &=&{\overline H},\nonumber\\
D^\lambda &=& {\overline D}-\lambda {\bf 1}_8, \nonumber\\
K^\lambda &=& {\overline K}-2\lambda t {\bf 1}_8, \nonumber\\
Q_I^\lambda &=&{\overline Q}_I,\quad\quad\quad\quad \quad\quad\quad\quad\quad\quad\quad\quad \quad\quad\quad\quad\quad\quad\quad\quad (I=1,2,3,4),\nonumber\\
{\widetilde Q}^\lambda_1&=&{\overline{\widetilde Q}}_1+2\lambda
(-E_{38}+E_{47}-E_{52}+E_{61}), \nonumber\\
{\widetilde Q}^\lambda_2&=&{\overline{\widetilde Q}}_2+2\lambda
(-E_{35}-E_{46}+E_{71}+E_{81}), \nonumber\\
{\widetilde Q}^\lambda_3&=&{\overline{\widetilde Q}}_3+2\lambda
(E_{36}-E_{45}-E_{72}+E_{81}), \nonumber\\
{\widetilde Q}^\lambda_4&=&{\overline{\widetilde Q}}_4+2\lambda
(E_{37}+E_{48}+E_{51}+E_{62}), \nonumber\\
S^\lambda_1&=&{\overline S}_1-2\lambda
(E_{12}-E_{21}+E_{34}-E_{43}+E_{56}-E_{65}+E_{78}-E_{87}),\nonumber\\
S^\lambda_2&=&{\overline S}_2,\nonumber\\
S^\lambda_3&=&{\overline S}_3,\nonumber\\
F^\lambda&=&4\lambda (E_{12}-E_{21}+E_{34}-E_{43}+E_{56}-E_{65}+E_{78}-E_{87}).
\eea
$F^\lambda$ enters the right hand side of the anticommutation relations
\bea\label{anticommf}
&\{Q^\lambda_3,{\widetilde Q}^\lambda_2\} =-\{Q^\lambda_2,{\widetilde Q}^\lambda_3\} =
S^\lambda_1+F^\lambda.&
\eea
The remaining (anti)commutation relations are the same as the corresponding ones of
$A(1,1)$
(in particular $S^\lambda_1$ appears in the right hand side of 
$\{Q^\lambda_4,{\widetilde Q}^\lambda_1\} =-\{Q^\lambda_1,{\widetilde Q}^\lambda_4\} =
S^\lambda_1$).

\subsection{The $(3,4,1)$ rep with $\lambda$ scaling dimension}

For $\lambda=0$, the $(3,4,1)$ $D$-module representation of the $A(1,1)$ superalgebra  is obtained by dressing
the $\lambda=0$ $D$-module representation of the root supermultiplet $(4,4)$. The dressing
operator is the diagonal $8\times 8$ matrix $S$, with diagonal elements 
$diag (S)=(1,1,1,\partial_t,1,1,1,1)$. Let $X$ be a generator of the $\lambda=0$
root $D$-module representation of $A(1,1)$. The corresponding generator ${\overline X}$ of the $(3,4,1)$
$D$-module representation is given by $
{\overline X}= S X S^{-1}$. 
For $\lambda\neq 0$ the $(3,4,1)$ $D$-module representation of the $D(2,1;\alpha)$ superalgebra is obtained. It is expressed by the supermatrices $X^\lambda$. In this case the relation between $\alpha$ and $\lambda$ is given by
\bea\label{alphalambda341}
\alpha&=& -\lambda.
\eea
We have
\bea\label{341}
H^\lambda &=&{\overline H},\nonumber\\
D^\lambda &=& {\overline D}-\lambda {\bf 1}_8,\nonumber\\
K^\lambda &=& {\overline K}-2\lambda t {\bf 1}_8,\nonumber\\
Q_I^\lambda &=&{\overline Q}_I,\quad\quad\quad\quad \quad\quad\quad\quad\quad\quad\quad\quad \quad\quad\quad\quad\quad\quad\quad\quad (I=1,2,3,4),\nonumber\\
{\widetilde Q}^\lambda_1&=&{\overline{\widetilde Q}}_1+2\lambda
(E_{47}-E_{52}+E_{61}-E_{83}), \nonumber\\
{\widetilde Q}^\lambda_2&=&{\overline{\widetilde Q}}_2+2\lambda
(-E_{46}-E_{53}+E_{71}+E_{82}), \nonumber\\
{\widetilde Q}^\lambda_3&=&{\overline{\widetilde Q}}_3+2\lambda
(-E_{36}+E_{63}-E_{72}+E_{81}), \nonumber\\
{\widetilde Q}^\lambda_4&=&{\overline{\widetilde Q}}_4+2\lambda
(E_{48}+E_{51}+E_{62}+E_{73}), \nonumber\\
S^\lambda_1&=&{\overline S}_1+2\lambda (-E_{12}+E_{21}-E_{56}+E_{65}) ,\nonumber\\
S^\lambda_2&=&{\overline S}_2+2\lambda (-E_{13}+E_{31}-E_{57}+E_{75}),\nonumber\\
S^\lambda_3&=&{\overline S}_3+2\lambda (E_{23}-E_{32}+E_{67}-E_{76}),\nonumber\\
W^\lambda_1&=&{\overline W}_1+2\lambda (2E_{12}-2E_{21}+E_{56}-E_{65}+E_{78}-E_{87}),\nonumber\\
W^\lambda_2&=&{\overline W}_2+2\lambda (2E_{13}-2E_{31}+E_{57}-E_{68}-E_{75}+E_{86}),\nonumber\\
W^\lambda_3&=&{\overline W}_3+2\lambda (-2E_{13}+2E_{31}-E_{58}-E_{67}+E_{76}+E_{85}).
\eea

\subsection{The inhomogeneous $(0,4,4)$ case}

For global supersymmetry the $(0,4,4)$ supermultiplet admits a inhomogeneous extension
which is compatible with the (\ref{sqm}) superalgebra. The addition of the $D$ and $K$ generators for the correct value  $\lambda=-\frac{1}{2}$ fails to produce a superconformal algebra. A larger superalgebra is obtained due to the fact that the commutators $[Q_i, S_j]$, for
$i\neq j$, produce extra odd generators besides the supercharges $Q_I$ and their superconformal partners ${\widetilde Q}_I$.

\subsection{The inhomogeneous $(1,4,3)$ case}

The $D$-module representation based on the inhomogeneous $(1,4,3)$ supermultiplet
$|m>$ (such that $<m|^T=(x, g_1,g_2,g_3,1;\psi_1,\psi_2,\psi_3,\psi_4)$ ) is expressed in terms of the left action 
on $|m>$ resulting from the $9\times 9$ supermatrices (with $9=(5|4)$ in the  ${\bf Z}_2$-grading
decomposition). A rotation of the global supersymmetries leaves us, without loss of generality, with a single inhomogeneous parameter $c$. We obtain a $D$-module representation of the $A(1,1)$ superconformal algebra. The homogeneous $(1,4,3)$ supermultiplet at $\lambda=-1$ is recovered in the limit $c\rightarrow 0$. We have 
\bea\label{143inhom}
H&=& (E_{11}+E_{22}+E_{33}+E_{44}+E_{66}+E_{77}+E_{88}+E_{99})\partial_t,\nonumber\\
D&=& -(E_{11}+E_{22}+E_{33}+E_{44}+E_{66}+E_{77}+E_{88}+E_{99})t\partial_t +\frac{1}{2}(2E_{11}+E_{66}+E_{77}+E_{88}+E_{99}),\nonumber\\
K&=& -(E_{11}+E_{22}+E_{33}+E_{44}+E_{66}+E_{77}+E_{88}+E_{99})t^2\partial_t +t(2E_{11}+E_{66}+E_{77}+E_{88}+E_{99}),\nonumber\\
Q_1&=& (-E_{26}-E_{39}+E_{48}+E_{71})\partial_t+E_{17}-E_{62}+E_{84}-E_{93}+cE_{85},\nonumber\\
Q_2&=& (E_{29}-E_{36}-E_{47}+E_{81})\partial_t+E_{18}-E_{63}-E_{74}+E_{92}-cE_{75},\nonumber\\
Q_3&=& (-E_{28}+E_{37}-E_{46}+E_{91})\partial_t+E_{19}-E_{64}+E_{73}-E_{82},\nonumber\\
Q_4&=& (E_{27}+E_{38}+E_{49}+E_{61})\partial_t+E_{16}+E_{72}+E_{83}+E_{94},\nonumber\\
{\widetilde Q}_1 &=&E_{71}(t\partial_t-2)+(-E_{26}-E_{39}+E_{48})(t\partial_t-1)+t(E_{17}-
E_{62}+E_{84}-E_{93}) -ctE_{85},\nonumber\\
{\widetilde Q}_2 &=&E_{81}(t\partial_t-2)+(E_{29}-E_{36}-E_{47})(t\partial_t-1)+t(E_{18}-
E_{63}-E_{74}+E_{92}) -ctE_{75},\nonumber\\
{\widetilde Q}_3 &=&E_{91}(t\partial_t-2)+(-E_{28}+E_{37}-E_{46})(t\partial_t-1)+t(-E_{19}+
-E_{64}+E_{73}-E_{82}),\nonumber\\
{\widetilde Q}_4 &=&E_{61}(t\partial_t-2)+(E_{27}+E_{38}+E_{49})(t\partial_t-1)+t(E_{16}-
E_{72}+E_{83}+E_{94}),\nonumber\\
S_1&=&2E_{34}-2E_{43}+E_{67}-E_{76}+E_{89}-E_{98}+cE_{35},\nonumber\\
S_2&=&-2E_{24}+2E_{42}+E_{68}-E_{79}-E_{86}+E_{97}-cE_{25},\nonumber\\
S_3&=&2E_{23}-2E_{32}+E_{69}+E_{78}-E_{87}-E_{96}.\nonumber\\
\eea
\section{$D$-module reps for ${\cal N}=8$ SCA: the $\lambda=\frac{1}{4}$
root rep and $D(4,1)$}

The minimal linear supermultiplets of the global ${\cal N}=8$-extended $1D$ supersymmetry contain $8$ bosonic and $8$ fermionic component fields. Without loss of generality \cite{{pt},{krt}} the $(8,8)$ root supermultiplet can be expressed in terms of the supercharges  and the Hamiltonian  given by
\bea\label{88}
&Q_i=\left(\begin{array}{cc}
0 &\gamma_i\\
-\gamma_i\partial_t&0
\end{array}\right),\quad
Q_8=\left(\begin{array}{cc}
0 &{\bf 1}_8\\
{\bf 1}_8\partial_t&0
\end{array}\right),\quad
H=\left(\begin{array}{cc}
{\bf 1}_8\partial_t &0\\
0&{\bf 1}_8\partial_t
\end{array}\right),&
\eea
where $\gamma_i$ ($i=1,2,\ldots,7$) are seven $8\times 8$ matrices, the generators of the Euclidean $Cl(0,7)$ Clifford algebra ($\gamma_i\gamma_j+\gamma_j\gamma_i=-2\delta_{ij}
{\bf 1}_8$).\par
As in the previous cases, the construction of a $D$-module representation for an ${\cal N}=8$ superconformal algebra requires the introduction of the bosonic 
generators $D,K$ which, together with $H$, close the  $sl(2)$ conformal subalgebra.
For the moment we can assume, as an Ansatz, such generators being expressed by the diagonal 
operators
\bea\label{sl2n8}
D&=&\left(\begin{array}{cc}
-{\bf 1}_8(t\partial_t+\lambda) &0\\
0&-{\bf 1}_8(t\partial_t+\lambda+\frac{1}{2})
\end{array}\right),\nonumber\\
K&=&\left(\begin{array}{cc}
-{\bf 1}_8(t^2\partial_t+2\lambda t) &0\\
0&-{\bf 1}_8(t^2\partial_t+(2\lambda +1)t)
\end{array}\right),
\eea
with conformal weights $\lambda$ for the bosonic fields and $\lambda+\frac{1}{2}$ for the fermionic fields.\par
The superconformal partners ${\widetilde Q}_i, {\widetilde Q}_8$ are introduced via the commutators
\bea\label{kq}
\relax [K, Q_i]= {\widetilde Q}_i, &&[K, Q_8] = {\widetilde Q}_8.
\eea
Explicitly,
\bea\label{tildeq}
{\widetilde Q}_i=\left(\begin{array}{cc}
0 &t\gamma_i\\
-\gamma_i(t\partial_t+2\lambda)&0
\end{array}\right),&&
{\widetilde Q}_8=\left(\begin{array}{cc}
0 &t {\bf 1}_8\\
{\bf 1}_8(t\partial_t+2\lambda)&0
\end{array}\right).
\eea 
The closure of the superalgebra requires the introduction of the new bosonic generators
$S_i$:
\bea\label{news}
&\{Q_8, {\widetilde Q}_i\}=-\{{\widetilde Q}_8, Q_i\}= S_i,&
\eea
where
\bea\label{svalue}
S_i &=& \left(\begin{array}{cc}
-2\lambda\gamma_i&0\\
0&(1-2\lambda)\gamma_i
\end{array}\right).
\eea
The requirement of being a superconformal algebra implies that no further odd generator
has to be added to its odd sector. This is not the case for $\lambda\neq \frac{1}{4}$, due to the commutation relations (for $i\neq j$)
\bea\label{lambdaconstraint}
\relax [S_i, Q_j] &=&
\left(\begin{array}{cc}
0 &\gamma_i\gamma_j(1-4\lambda)\\
(4\lambda-1)\gamma_i\gamma_j\partial_t&0
\end{array}\right).
\eea
One should note that in the quaternionic case (three $4\times 4$ matrices $\gamma_i$
generating the $Cl(0,3)$ Clifford algebra), since $\gamma_i\gamma_j\propto \epsilon_{ijk}\gamma_k$,
no extra odd generator is produced from the right hand side of the commutators corresponding to (\ref{lambdaconstraint}). This is why an ${\cal N}=4$ superconformal algebra is produced for any value of $\lambda$, while to get
an ${\cal N}=8$ superconformal algebra $\lambda$ has to be fixed to a special value.  For $\lambda\neq \frac{1}{4}$ a consistent ${\cal N}=8$ superalgebra, with a much larger set of even and odd generators, can be defined. This superalgebra is not, however, a superconformal algebra.
\par
For $\lambda=\frac{1}{4}$ the extra even generators $W_{[ij]}$ are obtained from the commutators
\bea\label{wij}
[S_i, S_j]&=&W_{[ij]},
\eea
with
 \bea\label{wijvalue}
W_{[ij]}&=&\frac{1}{2}\left(\begin{array}{cc}
\Sigma_{[ij]} &0\\
0&\Sigma_{[ij]}
\end{array}\right),
\eea
where $\Sigma_{[ij]}=\frac{1}{2}[\gamma_i,\gamma_j]$.\par
No extra generator (even or odd) is further required to consistently close the superalgebra.
The $W_{[ij]}$ generators close the $so(7)$ algebra. Taken together with the $S_i$ generators,
they close the $so(8)$ algebra. \par
The even sector is given by the generators ${\cal G}_0=\{H,D,K,S_i, W_{[ij]}\}$
which close the $sl(2)\oplus so(8)$ algebra. The odd sector ${\cal G}_1=\{Q_i, Q_8, {\widetilde Q}_i, {\widetilde Q}_8\}$ contains $16$ odd generators. A closer inspection of the (anti)commutation relations proves that, for $\lambda=\frac{1}{4}$, the construction originated by the global ${\cal N}=8$ $(8,8)$ root supermultiplet produces a $D$-module representation for the $D(4,1)$ superalgebra. \par
The extension of this analysis, namely which ${\cal N}=8$ superconformal algebra can be originated from the
inhomogeneous ${\cal N}=8$ global supermultiplets or from the homogeneous $(k, 8,8-k)$
supermultiplets with $k=1,2,\ldots, 7$,  is left for future investigations.

\section{On superconformal invariant actions from $D$-module reps}

The generators of the $D$-module representations act as graded Leibniz derivatives on functions of the component fields. Let $\varphi_A,\varphi_B$ be two component fields with scaling dimension
$\lambda_A,\lambda_B$, respectively. Then, the dilatation operator $D$ satisfies
\bea
D(\varphi_A\varphi_B)&=& (D\varphi_A)\varphi_B +\varphi_A(D\varphi_B)= -\frac{d}{dt}(\varphi_A\varphi_B) -\lambda_{A\cdot B} (\varphi_A\varphi_B),
\eea
where the scaling dimension of the composite field $\varphi_A\cdot\varphi_B$  is $\lambda_{A\cdot B}=\lambda_A+\lambda_B$.\par
Similarly, the conformal generator $K$ defined in (\ref{daction}) satisfies
\bea
K(\varphi_A\varphi_B)&=& (K\varphi_A)\varphi_B +\varphi_A(K\varphi_B)= -t^2\frac{d}{dt}(\varphi_A\varphi_B) -2\lambda_{A\cdot B} t(\varphi_A\varphi_B).
\eea
The action of the global supercharges $Q_I$ as graded Leibniz derivatives was the starting point
in \cite{{krt},{top1},{grt},{gkt}} to derive one-dimensional supersymmetric sigma-models from global supermultiplets. \par
For ${\cal N}=4$ a supersymmetric action, invariant by construction under the (\ref{sqm})
global superalgebra,  is
obtained for the $(k, {\cal N}, {\cal N}-k)$ supermultiplet  in terms of the Lagrangian
\bea\label{lagrangian}
{\cal L}&=& Q_4Q_3Q_2Q_1[F({\vec x})],
\eea
 where $F({\vec x})$ is an arbitrary function (named {\em the prepotential}) of the propagating bosons ${\vec x}$ (this construction requires the dimensionality $k$ of the target manifold to be $k>0$).\par
The Lagrangians recovered from (\ref{lagrangian}) have the correct dimensionality of a standard kinetic term, including the quadratic time-derivatives of the  propagating bosons. \par
For ${\cal N}=8$, in order to produce 
a supersymmetric sigma-model with the correct dimensionality of the standard kinetic term,
the above construction has to be enlarged. One possibility, which works in most of the cases
(for $k\geq 2$, see \cite{gkt}, where an alternative, more general picture is also described) consists in picking four out of the eight supercharges  ($Q_i$, for $i=1,2,3,4$) 
and use them as in (\ref{lagrangian}). Their supersymmetry is manifestly realized. The invariance under the remaining four supercharges ($Q_j$,
for $j=5,6,7,8$) is recovered by constraining their action on the Lagrangian to be a time-derivative: 
\bea\label{qjinv}
Q_j {\cal L}&=& \frac{d}{dt} P_j
\eea 
for some functions $P_j$ of the component fields and their time-derivatives.
As a result the prepotential $F({\vec x})$ is no longer unconstrained. Explicit computations prove that the
invariance under the global ${\cal N}=8$ supersymmetry is achieved if $F$  satisfies the harmonic condition\cite{{grt},{gkt}}, see also \cite{{bikl},{ils}}
\bea\label{harmonic}
\Box F &=& 0,
\eea
where ``$\Box$" stands for the $k$-dimensional Laplacian operator. \par
A superconformal mechanics is derived by constructing first a global supersymmetry via, for instance, the (\ref{lagrangian}) Lagrangian. Next, the Lagrangian is constrained by requiring the action of $K$ to produce a time-derivative:
\bea\label{kinv}
K{\cal L}&=&\frac{d}{dt} M,
\eea
for some function $M$ of the component fields and their time-derivatives.
\par
The invariance under the global supercharges $Q_I$ ($I=1,\ldots, {\cal N}$) and the conformal operator $K$ implies 
the invariance under the full superconformal algebra.  This is because the remaining generators in
the ${\cal N}=4$ and ${\cal N}=8$ superconformal algebras are obtained by repeatedly applying
the (anti)commutators among $K$ and the $Q_I$'s.\par
For ${\cal N}=8$ the extra (\ref{kinv}) constraint has to be added to the set of (\ref{qjinv}) 
constraints. \par
This general framework can be applied to derive a superconformal mechanics for
individual ${\cal N}$-extended supermultiplets or for a system of ${\cal N}$-extended supermultiplets in interaction (the prepotential $F$ in this case is a function of all propagating bosons, belonging to the different supermultiplets). We show explicitly how it works in some selected examples.\par
Let us take for simplicity an ${\cal N}=4$ $(1,4,3)$ global supermultiplet (its unique propagating boson is $x$) and apply (\ref{lagrangian}) to the prepotentials $F=x^\beta$ (with $\beta$ an arbitrary constant) and $F=x\ln x$, respectively. The scaling dimension $\lambda$ of the propagating boson $x$ (the overall conformal weight of the supermultiplet) is determined by requiring that the constraint (\ref{kinv}) should be satisfied. A straightforward computation proves that (\ref{kinv}) is satisfied provided that $\lambda =-\frac{1}{\beta}$ for the first Lagrangian
and $\lambda=-1$ for the second Lagrangian (in this second case the superconformal invariance
is recovered even in the presence of the inhomogeneous transformations depending on $c$, see 
subsection {\bf 5.6}). \par
Up to a total derivative the Lagrangian obtained from (\ref{lagrangian}) with $F=x^\beta$ is given by
\bea \label{xbetalagrangian}
{\cal L}&=& 
-\beta(\beta-1)x^{\beta-2}\left({\dot x}{\dot x}-\psi{\dot\psi}-
\psi_i{\dot\psi}_i+g_ig_i\right) -\nonumber\\
&&\beta (\beta-1)
(\beta-2)x^{\beta-3}\left(\psi\psi_ig_i+\frac{1}{2}\epsilon_{ijk}g_k\psi_j\psi_i \right)+\nonumber\\
&&
\frac{1}{6}\beta(\beta-1)(\beta-2)(\beta-3)x^{\beta-4}\epsilon_{ijk} \psi\psi_k\psi_j\psi_i.
\eea
Up to a total derivative the Lagrangian obtained from (\ref{lagrangian}) with $F= x\ln x$ and inhomogeneous $(1,4,3)$ supersymmetry transformations (without loss of generality the inhomogeneous
parameter $c\equiv c_3$ is taken along the third axis) is
\bea  \label{xlnxlagrangian}
{\cal L}&=& -\frac{1}{x}\left({\dot x}{\dot x}-\psi{\dot\psi}-
\psi_i{\dot\psi}_i+g_ig_i+g_3c\right) -\nonumber\\
&&\frac{1}{x^2}\left(-\psi\psi_ig_i-\psi\psi_3c+\frac{1}{2}\epsilon_{ijk}g_k\psi_j\psi_i \right)+\nonumber\\
&&
\frac{1}{3x^3}\epsilon_{ijk} \psi\psi_k\psi_j\psi_i.
\eea
Due to the relation $\lambda=-\frac{1}{\beta}$ and formula (\ref{alphalambda143}) the Lagrangian (\ref{xbetalagrangian})
defines a superconformal mechanics invariant under the $D(2,1;\alpha)$ superconformal algebra
for $\alpha=-\frac{1}{\beta}$. \par
For any value of the constant parameter $c$, the Lagrangian (\ref{xlnxlagrangian}) defines a superconformal mechanics invariant under the $A(1,1)$ superalgebra.\par
The resulting actions are scale-invariant and contain no dimensional parameter. In the first  case
we have, e.g., $S=\int dt Q_4Q_3Q_2Q_1(x^\beta)$, with $[S]= -1 +4\times \frac{1}{2}-\frac{1}{\beta}\times \beta= 0$.  \par
The inhomogeneous parameter $c$ is essential \cite{{ikp},{dprt},{lt}} in order to introduce Calogero-type terms in the superconformal action. The purely bosonic limit of (\ref{xlnxlagrangian}), after solving the
algebraic equations of motion of the auxiliary fields, gives ${\cal L}\sim \frac{1}{x}(-{\dot x}^2+\frac{c^2}{4})$. In terms of the new variable $y=\sqrt{x}$ we can write ${\cal L} = 4{\dot y}^2+\frac{c^2}{4y^2}$. This property extends to multiparticle superconformal mechanics (see \cite{{bgk},{gl0}}).

\subsection{An ${\cal N}=4$ case: $n_\pm$ inhomogeneous $(1,4,3)$ chiral supermultiplets}

We illustrate here as an example the computation of the superconformal invariance for several interacting supermultiplets. Let us take $n$ inhomogeneous ${\cal N}=4$ $(1,4,3)$ supermultiplets defined by the component fields  (at a given $I=1,\ldots, n$) $(x^I; \psi^I, \psi_i^I; g_i^I)$, with $i=1,2,3$. Let us further assume (see the discussion in Section {\bf 3}) that $n_+$ supermultiplets are of positive chirality $(I=1,\ldots, n_+)$ and $n_-=n-n_+$ supermultiplets are of negative chirality $(I=n_++1, \ldots, n)$. In terms of the global ${\cal N}=4$ supersymmetry the transformations of the component fields are therefore given by
\bea\label{nmulttransform}
&
\begin{array}{lllllll}
Q_4 x^I& =&\psi^I, &&Q_ix^I&=&\psi_i^I,\\
Q_4 \psi^I& =&{\dot x}^I, &&Q_i\psi^I&=&-g_i^I,\\
Q_4 \psi_j^I& =&g_j^I, &&Q_i\psi_j^I&=&\delta_{ij}{\dot x}_j^I+s_I\epsilon_{ijk}(
g_k^I+c_k^I),\\
Q_4 g_j^I& =&{\dot \psi}_j^I, &&Q_ig_j^I&=&-\delta_{ij}{\dot \psi}^I-s_I\epsilon_{ijk}{\dot \psi}_k^I.\\
\end{array}
&
\eea
The signs $s_I=\pm 1$ define the chirality of the corresponding supermultiplet.
The inhomogeneous parameters $c_k^I$ are arbitrary.\par
Up to a total derivative, the Lagrangian ${\cal L}=Q_4Q_3Q_2Q_1(F)$ derived from (\ref{lagrangian}) can be expressed as
\bea\label{nmultlagrangian}
{\cal L}&=&\sum_{IJ} s_IF_{IJ}\left(-{\dot x}^I{\dot x}^J+\psi^J{\dot\psi}^I+
\psi_i^J{\dot\psi}_i^I-g_i^J(g_i^I+c_i^I)\right)+\nonumber\\
&&\sum_{IJK}F_{IJK}\left( -s_I\psi^K\psi_i^J(g_i^I+c_i^I)+\frac{1}{2}\epsilon_{ijk}g_k^K\psi_j^J\psi_i^I )  \right)+\nonumber\\
&&\sum_{IJKL}F_{IJKL}\left(\frac{1}{6}\epsilon_{ijk} \psi^L\psi_k^K\psi_j^J\psi_i^I\right)
\eea
(here $F_I\equiv \frac{\partial}{\partial x_I}F$ and similarly for the higher order derivatives). The summation over $i,j,k$ is understood.\par
By inserting in (\ref{nmultlagrangian})  the prepotential 
\bea\label{Fsum}
F&=& \sum_{MN} A_{MN}~x_M\ln x_N,
\eea
where $A_{MN}$ is an arbitrary constant matrix, we obtain that the system of $n_\pm$ interacting
supermultiplets is superconformally invariant under the $A(1,1)$ superalgebra
(each supermultiplet has conformal weight $\lambda=-1$). The constraint (\ref{kinv}) induced by $K$ gets translated into the set of constraints, for $r=2,3,4$,
\bea\label{kconstraints}
x^MF_{MI_1\ldots I_r} = \gamma^{(r)} F_{I_1\ldots I_r}, &\quad&\gamma^{(r)}=1-r,
\eea
which are obviously satisfied if $F$ is given by (\ref{Fsum}).

\section{Conclusions and perspectives}

In this work we extended the minimal (homogeneous and inhomogeneous) linear supermultiplets of the one-dimensional ${\cal N}$-extended global supersymmetry algebra
({\ref{sqm}) to $D$-module representations of superconformal algebras. In the superconformal case the $D$-module representations are determined by the scaling dimension (conformal weight) $\lambda$ of the 
associated supermultiplet. \par
For the ${\cal N}=2$ and ${\cal N}=4$ extensions the analysis here presented
is exhaustive. We proved in particular that the ${\cal N}=4$ supermultiplets $(k, 4, 4-k)$
with $k=1,3,4$ induce, for $\lambda\neq 0$, $D$-module representations of the $D(2,1;\alpha)$
superconformal algebra. Different relations between $\alpha$ and $\lambda$ are found for each value of $k$, see formulas (\ref{alphalambda143}), (\ref{alphalambda341}) and (\ref{alphalambda44}). This result, combined with the mathematical property that different values of $\alpha$ related by an $S_3$ transformation define the same superalgebra $D(2,1;\alpha)$ (formula (\ref{equivalpha})), puts non-trivial restriction on superconformal model building. Several supermultiplets can be accommodated into a $D(2,1;\alpha)$ superconformal invariant model only if each supermultiplet separately carries
a representation of that superalgebra. \par
We also determined under which condition the global inhomogeneous supermultiplets induce
a $D$-module representation for an associated superconformal algebra. We mentioned that a
inhomogeneous parameter gives rise, in superconformally invariant actions, to the presence
of Calogero-type terms. \par
For ${\cal N}=8$ we proved that the $(8,8)$ ``root" supermultiplet of the global supersymmetry induces a $D$-module representation for the $D(4,1)$ superalgebra at
the specific value $\lambda=\frac{1}{4}$ of the scaling dimension. For $\lambda\neq \frac{1}{4}$
a consistent superalgebra is recovered. It is, however,  not an ${\cal N}=8$ superconformal algebra and possesses a large set of extra generators.\par
The construction here discussed uses only two inputs, the list of global supermultiplets and the
assumption that two extra diagonal generators $D$, $K$ have to be added. Together with $H$,
the Hamiltonian, they close an $sl(2)$ conformal subalgebra. $D$, $K$ are uniquely defined in terms of the scaling dimension $\lambda$.\par
The construction can be straighforwardly extended to more complicated cases. The next relevant question to be answered is the determination of the ${\cal N}=8$ superconformal algebras that can be recovered from
the homogeneous ${\cal N}=8$ global supermultiplets $(k, 8, 8-k)$ at a given scaling dimension $\lambda$, for $k=1,2,\ldots, 7$ (or from the inhomogeneous supermultiplets, with $k=1,2,3$).
This question will be answered with the help of a computer algebra package and presented in a 
forthcoming work.\par
The information contained in a $D$-module representation allows to construct superconformal mechanics in a Lagrangian setting. 
A general framework has been discussed in Section {\bf 7}. 
The global supercharges acting (as graded Leibniz derivatives)  on an arbitrary prepotential function $F({\vec x})$ of the propagating bosons allow to define a globally ${\cal N}=4$ supersymmetric
action. A constraint (\ref{kinv}) on the prepotential $F$ produced by the conformal generator $K$ is sufficient to
determine the full ${\cal N}=4$ superconformal invariance. As a specific example we presented the $A(1,1)$ superconformal invariant action of a system of interacting ${\cal N}=4$ $(1,4,3)$ inhomogeneous
supermultiplets of different chirality (the (\ref{nmultlagrangian}) Lagrangian with the (\ref{Fsum}) prepotential).\par
The generalization of this procedure to the construction of ${\cal N}=4$ superconformal mechanics for other sets of interacting supermultiplets (with the restriction mentioned before
on the allowed types of supermultiplets) is immediate.\par
The extension to the ${\cal N}=8$ superconformal mechanics requires a prepotential $F$ satisfying
the harmonic constraint, equation (\ref{harmonic}), besides the (\ref{kinv}) constraint produced by the
conformal generator $K$.\par
The determination of all $D$-module representations induced by global supermultiplets for
${\cal N}=4$ superconformal and (in the forthcoming paper) ${\cal N}=8$ superconformal algebras
is the starting point for a systematic investigation of superconformal mechanics in a Lagrangian setting. \par
This approach differs substantially from other investigations on superconformal mechanics based on superfields and/or a Poisson brackets structure.\par
The present method can be applied to solve open problems. Let us just mention one example among several possible choices. There exists a unique scale-invariant and globally ${\cal N}=8$ supersymmetric action defined for the $(2,8,6)$ inhomogeneous supermultiplet \cite{lt}. The scheme here discussed can be applied to determine whether this unique action is also superconformally invariant (under one of the ${\cal N}=8$ superconformal algebras).\par
There is no obstacle in extending this construction to the case of global ${\cal N}=9$ supercharges and searching for possible $D$-module representations of ${\cal N}=9$ superconformal algebras. For ${\cal N}=9$ two candidate superconformal algebras are $B(1,3)$
and $B(4,1)$ \cite{dictionary}. This case is particularly important because it corresponds to the
dimensional reduction of the ${\cal N}=4$ SuperYang-Mills theory, resulting in a $(9,16,7)$ supermultiplet with $9$ global supersymmetries realized off-shell \cite{top1}.\par
We should finally mention that the present results are propaedeutic to the construction of twisted superconformal symmetries. In \cite{bt} a twisted ${\cal N}=2$ superconformal algebra
with BRST-type odd generators was derived from the ordinary (untwist) ${\cal N}=2$ superconformal algebra  based on its $(1,2,1)$ $D$-module representation. The important program of twisting superconformal symmetries requires the preliminary construction of $D$-module representations for ${\cal N}=4$ superconformal algebras (whose complete list is here given), ${\cal N}=8$ superconformal algebras and so on.

~
\\ {~}~
\par {\large{\bf Acknowledgments}}
{}~\par{}~\par
We acknowledge useful discussions with O. Lechtenfeld. This work was supported by CNPq.

\end{document}